\documentclass[9pt,twocolumn]{article}
 \pdfoutput=1
\usepackage{graphicx}
\usepackage{epstopdf}
\usepackage[load-configurations=abbreviations]{siunitx}
\DeclareSIUnit\molar{\mole\per\cubic\deci\metre}
\DeclareSIUnit\Molar{\textsc{m}} % siunitx typesets physical units in a consistent manner
\usepackage{booktabs} % booktabs provides professional formatting commands for tables
\usepackage{amsmath} % amsmath provides extra maths symbols
\usepackage{textcomp} % textcomp provides extra text symbols (like a degrees celsius symbol)
\usepackage{tabularx}
\usepackage[labelformat=simple]{caption}
\usepackage{subfig}

\usepackage[english,plain]{fancyref}
\usepackage{url}
\usepackage{hyperref}

\usepackage{grffile}

\usepackage{nicefrac}

 \usepackage[sort&compress,numbers]{natbib}
\usepackage{color} 
\title{Colloidal Particles at Chiral Liquid Crystal Interfaces}
\author{Anne Claire Pawsey$^{1,2*}$, Juho Lintuvuori$^{1}$}
\date{January 2014}

\begin{document}
\maketitle
\noindent{\footnotesize{$^{1}$SUPA, School of Physics and Astronomy, University of Edinburgh, Mayfield Road, Edinburgh, EH9 3JZ, UK \\ $^{2}$ Current address Rowett Institute of Nutrition \& Health, University of Aberdeen, Greenburn Road, Bucksburn, Aberdeen, AB21 9SB\\
$^{*}$Email: a.pawsey@abdn.ac.uk}}

\begin{abstract}
Colloidal particles trapped at an interface between two fluids can form a wide range of different structures. Replacing one of the fluid with a liquid crystal increases the complexity of interactions and results in a greater range of possible structures. New behaviour emerges when colloidal particles interact with defects in the liquid crystal phases. Here we discuss the templating of colloids at a cholesteric isotropic interface. 
\end{abstract}

\section{Introduction}

Liquid crystal -- colloid composites are an exciting class of soft materials which possess tunable elastic and optical properties. In bulk samples of colloidal particles dispersed in liquid crystals, the colloids can assemble into a wide range of structures. In nematics, colloids can form chains~\cite{Loudet2000} and even a defect glass~\cite{Wood2011}. Adding a degree of twist by using a chiral (cholesteric) liquid crystal (CLC) increases the range of possibilities. The pitch length --- the distance in which the nematic director rotates by $2\pi$ --- of the cholesteric adds an additional  length scale to the system similar to the colloidal length scale. Competition between these mesoscopic scales leads to exciting new physics. Colloids form plates \cite{Hijnen2010}, regular arrays tied together with knotted defects~\cite{Tkalec2011} and three dimensional colloidal crystals ~\cite{Ravnik2011a}. The dynamics of the system also become highly non linear and non stokesian \cite{Lintuvuori2010,Lintuvuori2011}. These composite materials have potential applications for example as biosensors~\cite{Lin2011}. These applications involve contact with another fluid phase and hence the colloids interact with a fluid--fluid interface. 

The interface between two immiscible fluids is energetically expensive. The two fluids will seek to minimise the surface area in contact. Colloidal particles which are initially mixed into one of the two fluids can sequester to the interface and become trapped. A colloid removes a portion of the the interface and hence lowers the interfacial energy cost. In simple fluids, where the Young equation holds, the energy saving is expressed as 
\begin{equation}
\Delta G = -\pi r^{2} \gamma (1 \pm \mid\cos{\theta}\mid)^{2}, 
\end{equation}
where $\gamma$ is the interfacial tension, $\theta$ is the contact angle of the colloid with the interface  and $r$ is the particle radius~\cite{Binks2006}. For the system discussed below with interfacial tension $\gamma\sim$~\SI{4}{\milli\newton\per\metre}, a colloid radius of $r=$\SI{0.5}{\micro\metre} and assuming that the colloid is equally wetted by both fluids, $\theta=$\SI{90}{\degree}, the energy saving is $\sim 10^7k_BT$, thus thermal fluctuations will be unable to remove colloids from the interface. It is also important to note that in contrast to molecular surfactants, the colloids do not alter the interfacial tension between the liquids.

Colloidal particles  are commonly used to stabilise fluid--fluid interfaces. Careful control of the colloidal surface allows a variety of structures to be produced from simple particle stabilised emulsions~\cite{Binks2006} to more complex bicontinuous structures known as bijels \cite{Herzig2007}.  Conversely, fluid--fluid interfaces can be employed to create ordered arrangements of particles trapped on the interface ~\cite{Pieranski1980a, Stamou2000, Aveyard2000}. 

Particles at fluid -- fluid interfaces can self assemble at into a wide range of structures~\cite{Binks2006}  Deformations in the interface caused by gravity, shape anisotropy or surface roughness create interactions between particles. These capillary interactions, responsible for the clusters of cornflakes on the surface of the milk at breakfast \cite{Vella2005} are also an important tool in micro and nano-particle assembly. On flat interfaces between simple fluids, clusters of particles form at random locations \cite{Cavallaro2011}.  The location of clusters can be templated by manipulating the curvature of the interface~\cite{Cavallaro2011}. By replacing one of the fluids with a liquid crystal, colloid self--assembly can be templated even on a flat interface by making use of topological defects in the bulk \cite{Cavallaro2013} or at the interface between a simple fluid and the liquid crystal ~\cite{Mitov2002,Lintuvuori2013}. 

At simple fluid--fluid interfaces, the colloidal interactions which exist in the bulk phase are maintained. Ordered arrays of repulsive colloids are seen~\cite{Pieranski1980a}, some of which can be extraordinarily sparse~\cite{Horozov2003}. At higher surface coverage the arrangements become disordered~\cite{Pieranski1980a}. Additional interactions can be created by interfacial deformations. Such deformations can be created either via the weight of the colloid, the pinning of the contact line on the colloid or by distorting the whole interface. The interfacial deformations lead to capillary interactions which can be both attractive or repulsive~\cite{Oettel2008}.

The relative importance of gravity can be found by calculating the Bond number (\fref{eq:bond_number_c2}) the ratio between the gravitational forces and the surface tension. 
\begin{equation}
Bo=\frac{\Delta \rho g r^2}{\gamma}, \label{eq:bond_number_c2}
\end{equation}
where $\Delta \rho$ is the difference in density between the solid and the fluid, $r$ is the particle size and $\gamma$ is the surface tension between the two fluids. In cases where the Bond number is small ($Bo \ll 1$) gravitational effects may be discounted. However, even in the case of small Bond number the deformations caused by surface roughness ~\cite{Danov2010} or particle shape ~\cite{Cavallaro2011,Loudet2006} may still be present and lead to anisotropic interactions.

\section{Colloids at liquid crystal interfaces}
When one of the fluids is replace by a liquid crystal additional effects come into play due to the anisotropic nature of the LC. Colloids alter the ordientation of the director at their surface. The strength of this influence is characterised by an anchoring strength $W$. For finite anchoring ($W > 0$) this imposed alignment breaks the orientational order of the liquid crystal and and leads to the formation of defects. The  type of defect created depends on the anchoring type and strength at the colloid surface and the colloid size \cite{Stark2001a}.  At the fluid--LC interface the defects remain in the LC phase but are altered as the colloid is only partially submerged in the LC \cite{Gharbi2011a, Koenig2010}. Altering the anchoring at the fluid--LC interface by adding a surfactant to the fluid phase has been shown to alter the defect structure at the colloid and hence the interactions between colloids \cite{Koenig2010}.

Recent work by Cavallaro {\it{et al..}} \cite{Cavallaro2013} has shown that colloids at a nematic--air interface interact with defects in the bulk of the liquid crystal. In equilibrium nematics are defect free so defects were created by using a micro-pillar array. The pillars were surrounded by ring defects at their mid-point were in the bulk of the LC. The defects set up a distorted director field which templated the assembly of the colloids into a ring on the interface above the defect.

The use of defects to template colloidal assembly can be extended to using pre-existing defects as templates. Defects inherent in nematic droplets can have been shown to localise colloidal particles ~\cite{Mondiot2013, Whitmer2013}. In  bulk samples of blue phases and twist grain boundary phases the defects are also preferential sites for colloids \cite{Cordoyiannis2013}. Nano-particles have been shown to assemble at the interface of a CLC in distinct stripes related to the cholesteric helix~\cite{Mitov2002,Bitar2011}.  In this article the use of defected regions inherent in the cholesteric fingerprint texture as a templates for assembly of micron--sized particles is discussed.
 
The interaction between colloids and liquid crystal defects is rarely straight forward. Colloidal particles themselves produce defects in LC phases. How the defects associated with a colloidal particle interact with other defects present in the system can have a profound influence on the success or otherwise of templating.  

\section{Colloids at cholesteric interfaces}
\begin{figure}[h]
\centering
\subfloat[]{\includegraphics[width=0.8\columnwidth]{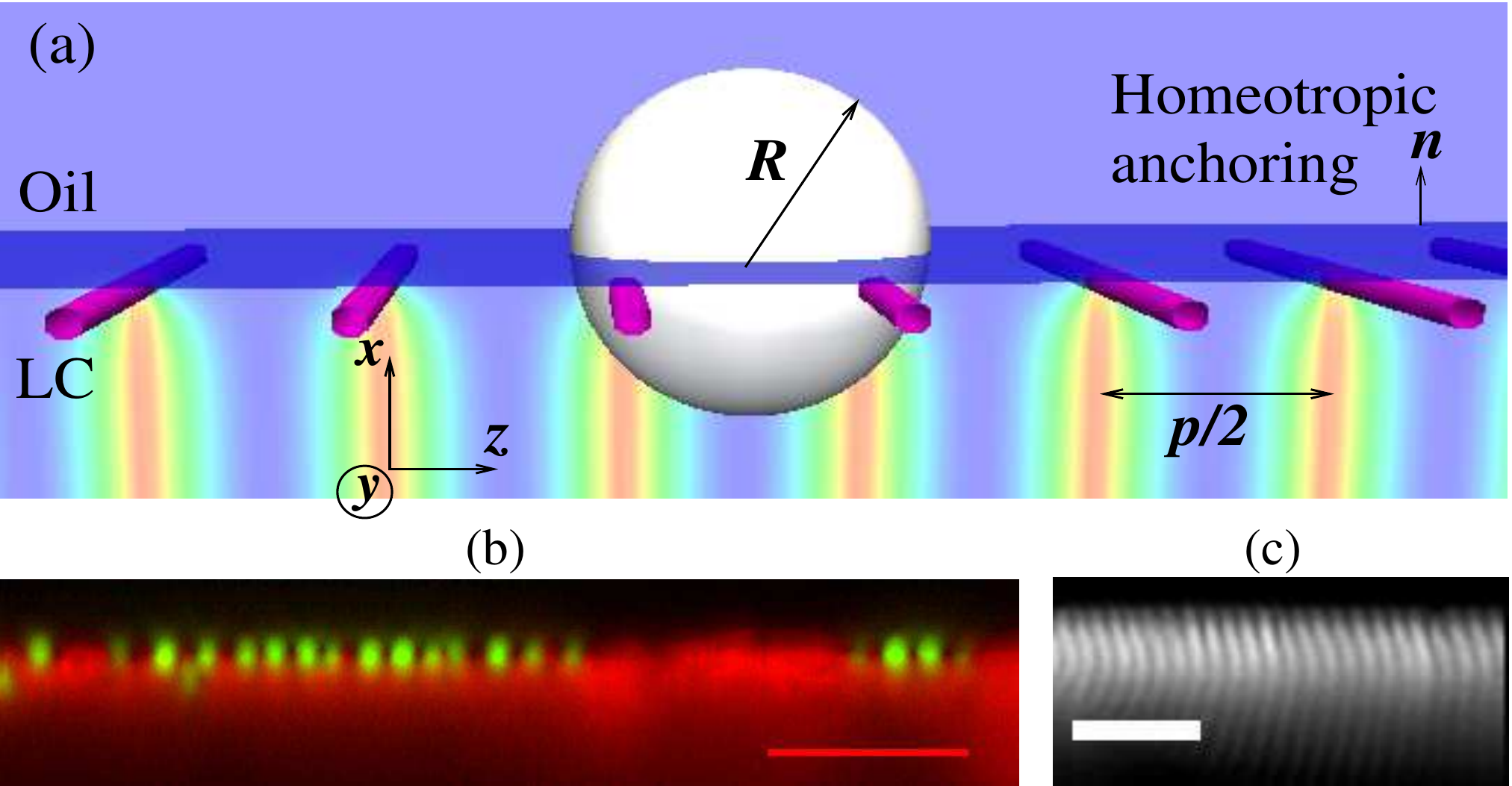}}
\caption[The interface viewed in cross section]{The interface viewed in cross section. (a) Schematic of the geometry: one particle of radius R is trapped on the interface between a cholesteric LC and an isotropic fluid (an immiscible oil). In the LC phase, red and blue correspond to director field along y and x, respectively, pink lines are the defects. (b) Fluorescence confocal image of \SI{3}{\micro\metre} particles (green) trapped at the interface between cholesteric (red) and oil (dark) (scale bar \SI{20}{\micro\metre}). (c) Fluorescence confocal microscopy image of the cholesteric-oil boundary in the absence of particles, showing that the interface is flat (scale bar \SI{10}{\micro\metre}) Reproduced with permission from \cite{Lintuvuori2013} Copyright (2013) by The American Physical Society}\label{fig:int_schematic}
\end{figure}

Cholesteric liquid crystals possess orientational order but additionally break translational symmetry due to their chiral nature. The average orientation of the molecules, denoted by the director, $\mathbf{n}$, maps out a helix along an axis, $\vec{h}$. The distance taken for the director to rotate through $2\pi$ gives the material an intrinsic length scale, the pitch length, $p$. This periodic structure can be distorted in order to satisfy boundary conditions at a surface or in response to external fields or inclusions such as colloids. The layers can either be distorted elastically where the layers bend smoothly or via the formation of defects~\cite{Lubensky1972,Smalyukh2002}. 
\begin{figure}[t]
\centering
\includegraphics[width=0.6\columnwidth]{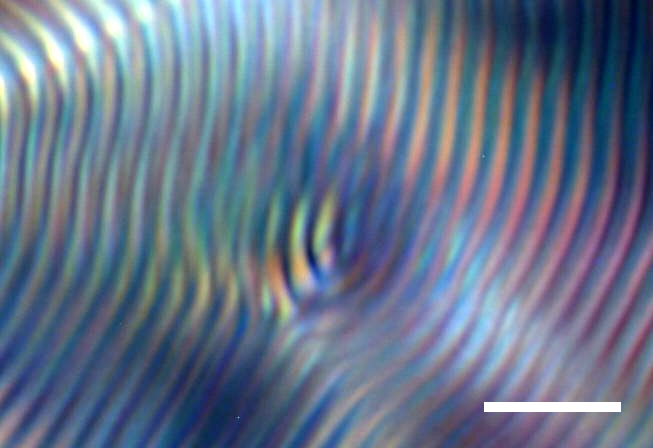}
\caption{The fingerprint texture at a cholesteric silicone oil interface viewed through the overlying silicone oil. The arrow indicates the helical axis direction. A $\chi$ edge defect can be seen in the centre of the image. Scale bar \SI{5}{\micro\metre}}\label{fig:fingerprint}
\end{figure}
In cholesterics the competition between the helical twist and surface anchoring of the liquid crystals leads to the appearance of striking textures. Where the anchoring is homeotropic (normal to the interface) at both top and bottom surfaces the helical axis is forced to lie parallel to these surfaces and the fingerprint texture is observed. This texture appears as a periodic array of bright and dark stripes as in \fref{fig:fingerprint} when imaged using polarising optical microscopy. Using a confocal microscope the periodic nature of the helix can also be seen in a vertical slice through the interface \fref{fig:int_schematic} (c). The fingerprint texture is an example of frustrated orientational order  as the constraints of both surface anchoring and helical twist cannot be satisfied simultaneously~\cite{Gennes1993}. For a reasonably high interfacial tension the frustration is resolved by the creation of a 2d array of defect lines at the interface, these are sketched in pink in \fref{fig:int_schematic} (a). 

\begin{figure}[h]
\centering
\subfloat[]{\includegraphics[width=0.45\columnwidth]{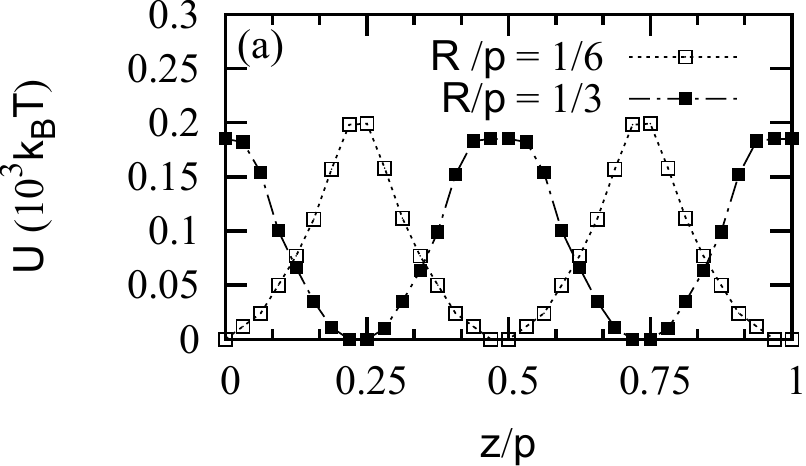}\label{fig:zero_anchoring_graph}}~
\subfloat[]{\includegraphics[width=0.45\columnwidth]{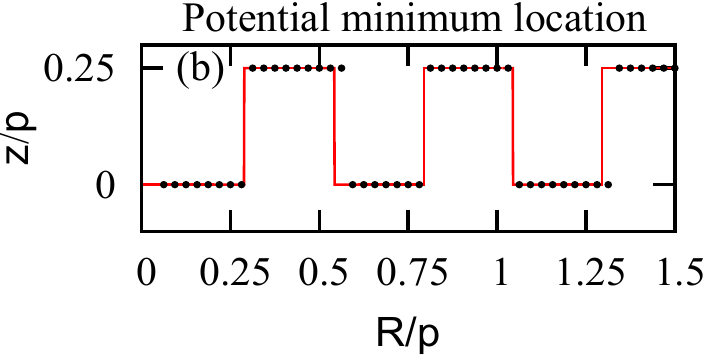}\label{fig:min_loc}}\\
\subfloat[]{\includegraphics[trim= 40mm 50mm 40mm 20mm, clip, width=0.5\columnwidth]{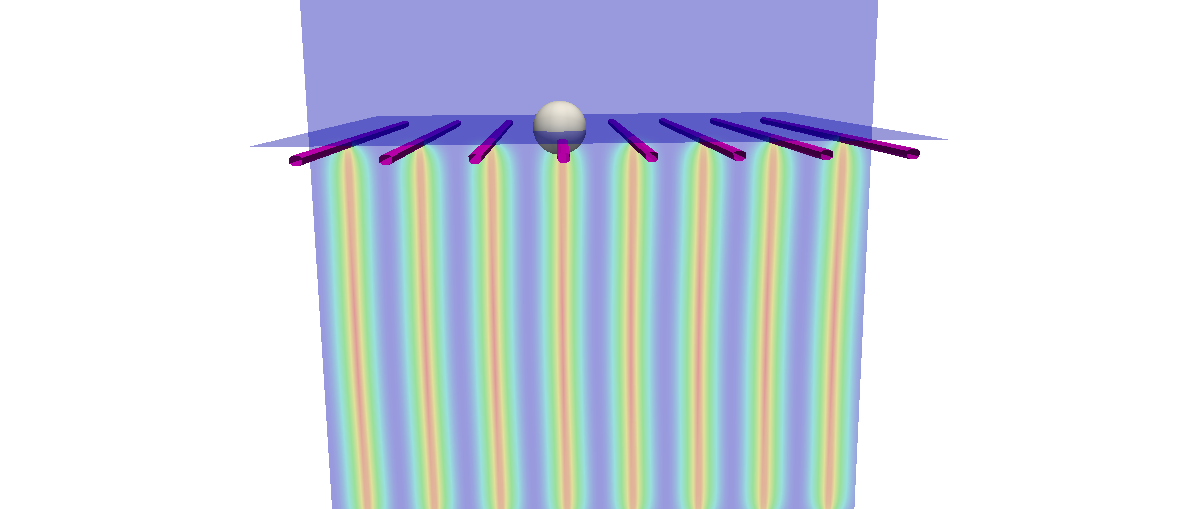}\label{fig:small_top}}~
\subfloat[]{\includegraphics[trim= 40mm 50mm 40mm 20mm, clip, width=0.5\columnwidth]{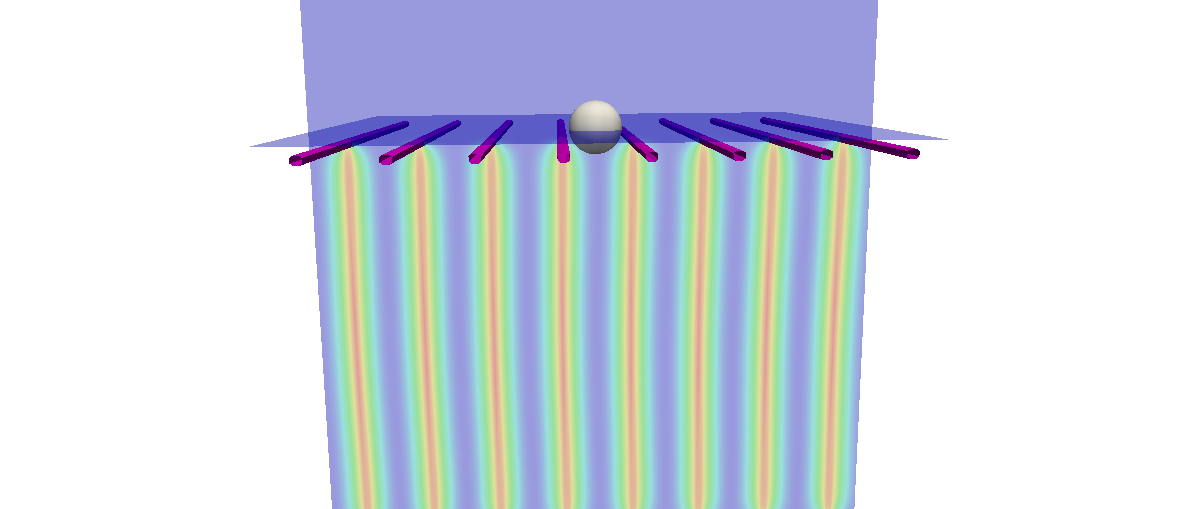}\label{fig:small_between}}\\ 
\subfloat[]{\includegraphics[trim= 30mm 50mm 30mm 20mm, clip, width=0.5\columnwidth]{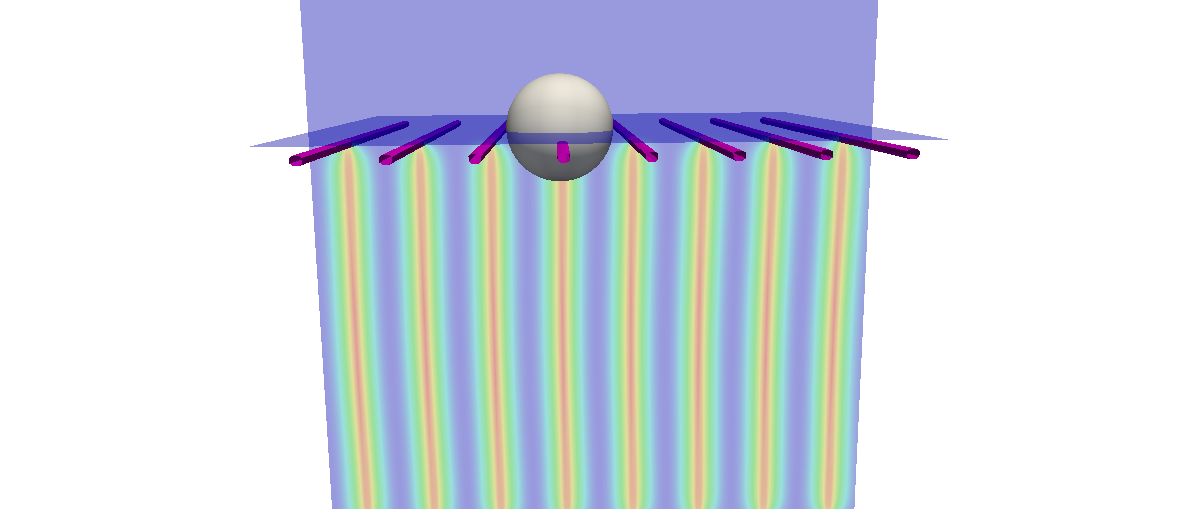}\label{fig:large_top}}~ 
\subfloat[]{\includegraphics[trim= 30mm 50mm 30mm 20mm, clip, width=0.5\columnwidth]{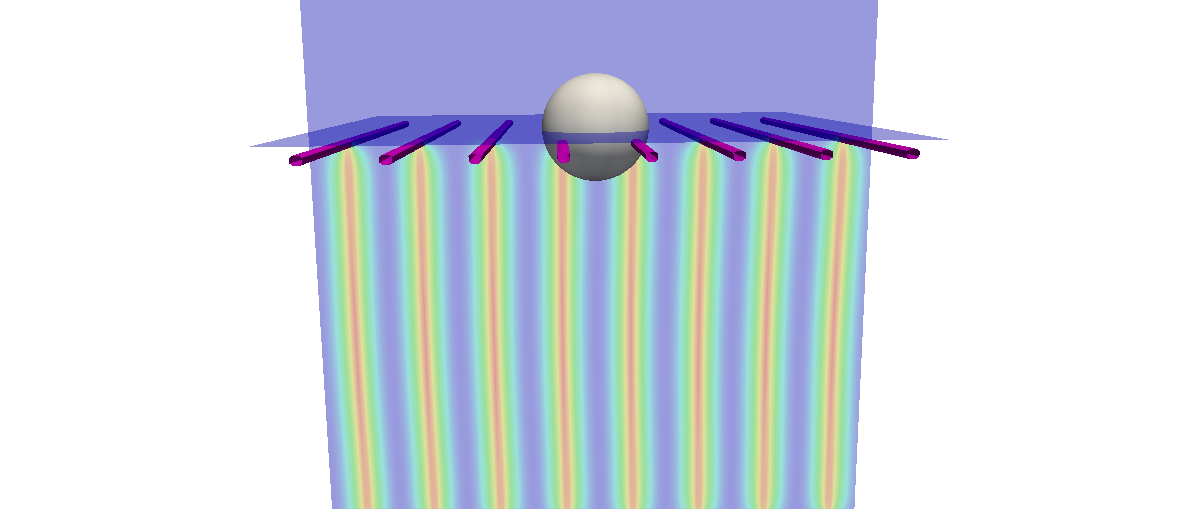}\label{fig:large_between}} 
\caption[Graphs and schematics of colloidal templating at the interface]{\protect\subref*{fig:zero_anchoring_graph} Plot of free energy of the colloid-LC system when the position of the particle is varied on the interface, for a colloid with zero anchoring ($WR/K=0$). z is the position from the defect core, p is the liquid crystal pitch. As the liquid crystal director field is head-tail symmetric the plot is symmetric both around a distance of 0.5 p and 0.25 p. \protect\subref*{fig:min_loc}  Plot of the distance from the disclination core of the minimum of the free energy (in units of  the pitch length p) as a function of $R/p$. The solid (red) line is a prediction based on the colloid covering the largest possible length  of disclination.  \protect\subref*{fig:zero_anchoring_graph} and \protect\subref*{fig:min_loc} from  \cite{Lintuvuori2013}.   \protect\subref*{fig:small_top} a small colloid $R/p=1/6$ centred on a defect will remove a section of defect, reducing the energy cost. \protect\subref*{fig:small_between} a small colloid between the defects will not remove any defect core. \protect\subref*{fig:large_top} a large colloid $R/p=1/3$ centred on a defect only removes that defect. \protect\subref*{fig:large_between} a large colloid between defects removes two defect cores. Adapted with permission from \cite{Lintuvuori2013} Copyright (2013) by The American Physical Society}\label{fig:zero_anchoring}
\end{figure}

The defect array costs the system additional elastic energy. Colloids trapped at the interface can reduce the energy of the system by removing a portion of the defected region. The results of simulations~\cite{Lintuvuori2013} of the colloid-LC system for colloids with zero anchoring strength $W = 0$ (colloids do not influence the alignment of the LC at their surfaces) in figures \ref{fig:zero_anchoring_graph} and \ref{fig:min_loc} show that the energy of the system  depends on the position of the colloid with respect to the defect lines. The most favourable position varies with the particle size to pitch ratio $R/p$.  The equilibrium position of the colloid's centre changes according to where the colloid covers the greatest total length of defect. Simulations in ~\cite{Lintuvuori2013} for $W\rightarrow 0$ predict a favourable position for the colloid centre either exactly on top of an interfacial defect or exactly half way between them as seen in \fref{fig:min_loc}.

\begin{figure}[h]
\centering
\subfloat[]{\includegraphics[width=0.5\columnwidth]{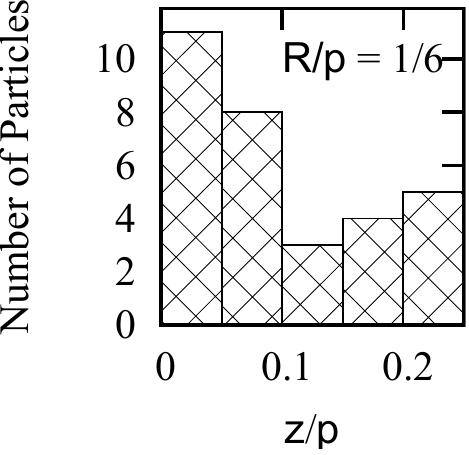}\label{fig:p_hist16}}~
\subfloat[]{\includegraphics[width=0.5\columnwidth]{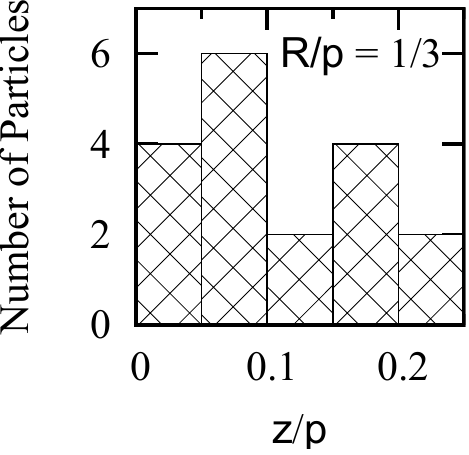}\label{fig:p_hist13}}\\
\subfloat[]{\includegraphics[trim=0mm 0mm 0mm 10mm,clip, width=0.5\columnwidth]{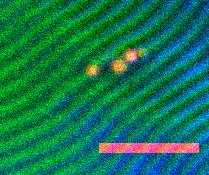}\label{fig:longP_chain}}~
\subfloat[]{\includegraphics[width=0.5\columnwidth]{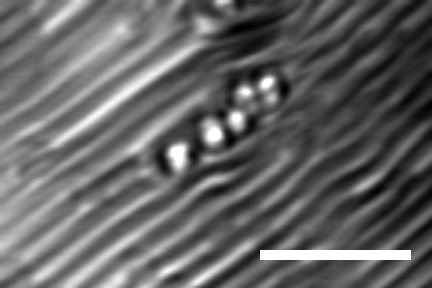}\label{fig:shortpchain}}
\caption[Positions of particles relative to interfacial stripes]{Experimental histograms of the distribution of isolated particles at the cholesteric-oil interface, for \protect\subref*{fig:p_hist16}  $\frac{R}{p}= \frac{1}{6}$ (R = \SI{0.5}{\micro\metre} and p = \SI{3.2}{\micro\metre}), and \protect\subref*{fig:p_hist13} $\frac{R}{p}= \frac{1}{3}$ ( R = \SI{0.5}{\micro\metre} and p = \SI{1.5}{\micro\metre}) \protect\subref*{fig:longP_chain} Fluorescence microscopy image showing an isolated particle and a dimer, each lying atop disclination lines $\frac{R}{p}= \frac{1}{6}$  \protect\subref*{fig:shortpchain} Transmission microscopy image of a longer chainlike aggregate $\frac{R}{p}= \frac{1}{3}$;  the disclination lattice is partially disrupted in this case Reproduced with permission from \cite{Lintuvuori2013} Copyright (2013) by The American Physical Society. Scale bars \SI{10}{\micro\metre}}\label{particle_hists}
\end{figure}

In experiments performed with melamine colloids trapped at a cholesteric silicone oil interface~\cite{Pawsey2012} the position of the particles with respect to the stripe texture was measured.  Histograms of particle position with respect to the defect are plotted in \fref{fig:p_hist16} and \ref{fig:p_hist13}. This distance is in units of $z/p$ where $z/p=0$ is the defect. For the particles in the longer pitch cholesteric $R/p=1/6$ shown in \fref{fig:p_hist16} there is evidence of templating. Colloids are found preferentially in localised in the defected regions. For the  particles in the shorter pitch mixture $R/p=1/3$ no clear templating effect is seen. 

The templating of the smaller particles at the interface suggests that the interfacial defects have control the colloid position. However, the lack of a preferred position for the $R/p =1/3$ samples indicates that the physics is more complicated than the  model discussed above. In the simple  model, the anchoring strength $W$ at the colloid was set to zero; the colloid acts as a free surface for the liquid crystal. Hence, the colloids had no effect on the alignment of the liquid crystal. 

\begin{figure}[h]
\centering
\subfloat[]{\includegraphics[width=0.25\columnwidth]{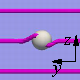}\label{fig:rewire1}}~
\subfloat[]{\includegraphics[width=0.25\columnwidth]{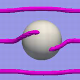}\label{fig:rewire2}}~
\subfloat[]{\includegraphics[width=0.25\columnwidth]{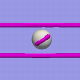}\label{fig:rewire3}}~
\subfloat[]{\includegraphics[width=0.25\columnwidth]{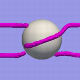}\label{fig:rewire4}}\\
\subfloat[]{\includegraphics[width=0.5\columnwidth]{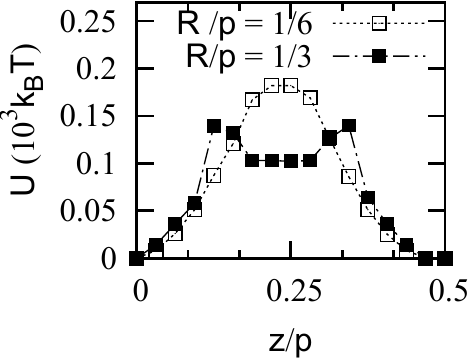}\label{fig:finite_w_1}}~
\subfloat[]{\includegraphics[width=0.5\columnwidth]{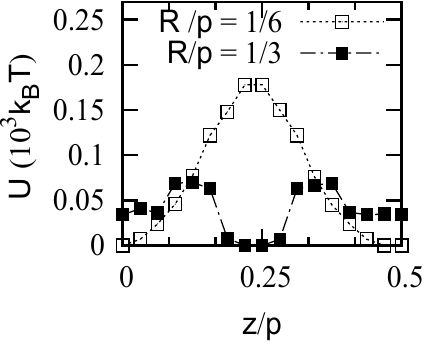}\label{fig:vary_alpha}}
\caption[Defect rewiring]{\protect\subref*{fig:rewire1} to \protect\subref*{fig:rewire4} Disclination structures found in simulations for a colloid trapped at the cholesteric-oil interface with strong anchoring at the particle surface. Particle size is varied at fixed surface anchoring W, and pitch p. The dimensionless ration $WR/K = 5.3$  \protect\subref*{fig:rewire1} and \protect\subref*{fig:rewire3}, and $WR/K = 10.67$ in \protect\subref*{fig:rewire2} and \protect\subref*{fig:rewire4}. In \protect\subref*{fig:rewire1} and \protect\subref*{fig:rewire2} the colloid is centered on a disclination, in \protect\subref*{fig:rewire3}and \protect\subref*{fig:rewire4} midway between disclinations. \protect\subref*{fig:finite_w_1} Plots of the free energy landscapes corresponding to the schematics.  \protect\subref*{fig:vary_alpha} at an intermediate anchoring strength $WR/K = 2/7$ the energy landscape becomes more complex and flatter  particularly in the case of the  ``larger'' ($R/p = 1/3$) particles. Adapted with permission from \cite{Lintuvuori2013} Copyright (2013) by The American Physical Society }\label{fig:rewire}
\end{figure}

In the experiments the anchoring strength is unknown but it is unlikely to be zero. Further simulations  at finite anchoring strength reveal that the colloids themselves create defects. These defects can interact with the interfacial defects as seen in \fref{fig:rewire1} to \ref{fig:rewire4}. For  $R/p =1/6$ this rewiring of the defects has very little effect on the shape of the free energy landscape \fref{fig:finite_w_1} (open symbols). However, for the $R/p =1/3$  case the energy landscape has been significantly altered. The position of the minimum has moved from its position of $z/p =0.25$ in the zero anchoring case to  $z/p =0$. This implies that the physics of the system is governed by anchoring strength $W$ as well as the particle size to pitch ratio $R/p$. 

In order to make comparisons a dimensionless ration $w=WR/K$ is defined. It is the strength of the anchoring at the colloidal surface relative to bulk elasticity in the liquid crystal. For small $w$ the colloid position depends solely on the ratio $R/p$ and seeks to cover the largest possible length of defect. As $w$ is increased the energy landscape becomes more complex.  For intermediate values of $w$  and $R/p=1/3$ the landscape is fairly flat and rough as seen in \fref{fig:vary_alpha}. This landscape is unlikely to be able to act as a good template for colloids. Assuming an intermediate anchoring strength $W$, the lack of templating in our experimental system could be accounted for by this complex landscape.

The sensitivity to the magnitude of $WR/K$ is seen in other systems where colloids interact with defects. For example, the size of nano particles~\cite{Dierking2012} and their surface anchoring~\cite{Rozic2011} has been shown to have a profound effect on their ability to stabilise blue phases.

\section{Conclusions}
Elasticity  mediated interactions have shown great promise for colloidal self assembly at both flat fluid--LC interfaces~\cite{Cavallaro2013} and on the surface of nematic droplets~\cite{Whitmer2013}. Here, we have shown that the by conflict between alignment at a CLC--isotropic oil interface and the helical nature of the CLC creates an array of interfacial defects. These defected regions  can act as templates for  the self assembly of colloidal particles. The success of the templating mechanism is sensitive to the balance between the anchoring strength at the colloid surface and the elasticity of the liquid crystal, $w=WR/K$. Understanding this interplay between colloids and defects suggests mechanisms by which LC colloid composites for example colloid stabilised blue phases could be tuned for greatest effect. 
\phantomsection

\section{Acknowledgments}
This work was funded by EPSRC Grants No. EP/ I030298/1, No. EP/E030173, and No. EP/E045316. A. C. P. was funded by the EPSRC Scottish Doctoral Training Centre in Condensed Matter Physics.

% \bibliographystyle{lctoday}
% \bibliogrAphy{draft1bib}


\begin{thebibliography}{10}

\bibitem{Loudet2000}
Loudet, JC, Barois, P, Poulin, P, 2000;\ {Colloidal ordering from
  phase separation in a liquid-crystalline continuous phase}. Nature, 407,
  6804:611--3.

\bibitem{Wood2011}
Wood, TA, Lintuvuori, JS, Schofield, AB, Marenduzzo, D, Poon, WCK,
  2011;\ {A Self-Quenched Defect Glass in a Colloid-Nematic Liquid
  Crystal Composite}. Science, 334, 6052:79--83.

\bibitem{Hijnen2010}
Hijnen, N, Wood, TA, Wilson, D, Clegg, PS, 2010;\ {Self-organization
  of particles with planar surface anchoring in a cholesteric liquid crystal.}
  Langmuir, 26, 16:13502--10.

\bibitem{Tkalec2011}
Tkalec, U, Ravnik, M, Copar, S, \v{Z}umer, S, Musevic, I,
  2011;\ {Reconfigurable Knots and Links in Chiral Nematic Colloids}.
  Science, 333, 6038:62--65.

\bibitem{Ravnik2011a}
Ravnik, M, Alexander, GP, Yeomans, JM, \v{Z}umer, S,
  2011;\ {Three-dimensional colloidal crystals in liquid crystalline
  blue phases.} Proc Natl Acad Sci USA, 108, 13:5188--92.

\bibitem{Lintuvuori2010}
Lintuvuori, JS, Marenduzzo, D, Stratford, K, Cates, ME,
  2010;\ {Colloids in liquid crystals: a lattice Boltzmann study}. J
  Mater Chem, 20, 46:10547.

\bibitem{Lintuvuori2011}
Lintuvuori, J, Stratford, K, Cates, M, Marenduzzo, D,
  2011;\ {Self-Assembly and Nonlinear Dynamics of Dimeric Colloidal
  Rotors in Cholesterics}. Phys Rev Lett, 107, 26:267802.

\bibitem{Lin2011}
Lin, IH, Miller, DS, Bertics, PJ, Murphy, CJ, de~Pablo, JJ, abbott, NL,
  2011;\ {Endotoxin-induced structural transformations in liquid
  crystalline droplets.} Science, 332, 6035:1297--300.

\bibitem{Binks2006}
Binks, B, Horozov, TS, 2006;\ \emph{{Colloidal particles at liquid
  interfaces}}, Cambridge University Press.

\bibitem{Herzig2007}
Herzig, EM, White, KA, Schofield, AB, Poon, WCK, Clegg, PS,
  2007;\ {Bicontinuous emulsions stabilized solely by colloidal
  particles.} Nature materials, 6, 12:966--71.

\bibitem{Pieranski1980a}
Pieranski, P, 1980;\ {Two Dimensional Colloidal Crystals}. Phys Rev
  Lett, 45, 7:569--572.

\bibitem{Stamou2000}
Stamou, D, Duschl, C, Johannsmann, D, 2000;\ {Long-range attraction
  between colloidal spheres at the air-water interface: the consequence of an
  irregular meniscus}. Phys Rev E, 62:5263--72.

\bibitem{Aveyard2000}
Aveyard, R, Clint, JH, Nees, D, Paunov, VN, 2000;\ {Compression and
  Structure of Monolayers of Charged Latex Particles at air / Water and Octane
  / Water Interfaces}. Langmuir, 16, 13:1969--1979.

\bibitem{Vella2005}
Vella, D, Mahadevan, L, 2005;\ {The ``Cheerios effect''}. American
  Journal of Physics, 73, 9:817.

\bibitem{Cavallaro2011}
Cavallaro, M, Botto, L, Lewandowski, EP, Wang, M, Stebe, KJ,
  2011;\ {Curvature-driven capillary migration and assembly of
  rod-like particles.} Proc Natl Acad Sci USA, 108, 52:20923--8.

\bibitem{Cavallaro2013}
Cavallaro, M, Gharbi, Ma, Beller, Da, \v{C}opar, S, Shi, Z, Baumgart, T, Yang,
  S, Kamien, RD, Stebe, KJ, 2013;\ {Exploiting imperfections in the
  bulk to direct assembly of surface colloids.} Proc Natl Acad Sci USA, 110,
  47:18804--8.

\bibitem{Mitov2002}
Mitov, M, Portet, C, Bourgerette, C, Snoeck, E, Verelst, M,
  2002;\ {Long-range structuring of nanoparticles by mimicry of a
  cholesteric liquid crystal.} Nature materials, 1, 4:229--31.

\bibitem{Lintuvuori2013}
Lintuvuori, JS, Pawsey, AC, Stratford, K, Cates, ME, Clegg, PS, Marenduzzo, D,
  2013;\ {Colloidal templating at a cholesteric-Oil interface:
  assembly guided by an array of disclination lines}. Phys Rev Lett, 110,
  18:187801.

\bibitem{Horozov2003}
Horozov, TS, Aveyard, R, Clint, JH, Binks, BP,
  2003;\ {Order--disorder transition in monolayers of modified
  monodisperse silica particles at the octane--water interface}. Langmuir,
  19, 7:2822--2829.

\bibitem{Oettel2008}
Oettel, M, Dietrich, S, 2008;\ {Colloidal interactions at fluid
  interfaces.} Langmuir, 24, 4:1425--41.

\bibitem{Danov2010}
Danov, KD, Kralchevsky, PA, 2010;\ {Capillary forces between
  particles at a liquid interface: general theoretical approach and
  interactions between capillary multipoles.} adv Colloid Interface Sci, 154,
  1-2:91--103.

\bibitem{Loudet2006}
Loudet, J, Yodh, A, Pouligny, B, 2006;\ {Wetting and Contact Lines of
  Micrometer-Sized Ellipsoids}. Phys Rev Lett, 97, 1:018304.

\bibitem{Stark2001a}
Stark, H, 2001;\ {Physics of colloidal dispersions in nematic liquid
  crystals}. Phys Rep, 351, 6:387--474.

\bibitem{Gharbi2011a}
Gharbi, MA, Nobili, M, In, M, Pr\'{e}vot, G, Galatola, P, Fournier, JB, Blanc,
  C, 2011;\ {Behavior of colloidal particles at a nematic liquid
  crystal interface†}. Soft Matter, 7, 4:1467.

\bibitem{Koenig2010}
Koenig, GM, Lin, IH, Abbott, NL, 2010;\ {Chemoresponsive assemblies
  of microparticles at liquid crystalline interfaces.} Proc Natl Acad Sci U SA, 107, 9:3998--4003.

\bibitem{Mondiot2013}
Mondiot, F, Wang, X, de~Pablo, JJ, Abbott, NL, 2013;\ {Liquid
  crystal-based emulsions for synthesis of spherical and non-spherical
  particles with chemical patches.} J am Chem Soc:2--5.

\bibitem{Whitmer2013}
Whitmer, JK, Wang, X, Mondiot, F, Miller, DS, Abbott, NL, de~Pablo, JJ,
  2013;\ {Nematic-field-driven positioning of particles in liquid
  crystal droplets}. Phys Rev Lett, 111, 22:227801.

\bibitem{Cordoyiannis2013}
Cordoyiannis, G, {Rao Jampani}, VS, Kralj, S, Dhara, S, Tzitzios, V, Basina, G,
  Nounesis, G, Kutnjak, Z, {Pati Tripathi}, CS, Losada-P\'{e}rez, P, Jesenek,
  D, Glorieux, C, Mu\v{s}evi\v{c}, I, Zidan\v{s}ek, A, ameinitsch, H, Thoen, J,
  2013;\ {Different modulated structures of topological defects
  stabilized by adaptive targeting nanoparticles}. Soft Matter, 9, 15:3956.

\bibitem{Bitar2011}
Agez, G, Bitar, R, Mitov, M, 2011;\ {Cholesteric liquid crystal
  self-organization of gold nanoparticles}. Soft Matter, 7, 6:8198--8206.

\bibitem{Lubensky1972}
Lubensky, T, 1972;\ {Hydrodynamics of cholesteric liquid crystals}.
  Phys Rev a, 6, JULY:452.

\bibitem{Smalyukh2002}
Smalyukh, I, Lavrentovich, O, 2002;\ {Three-dimensional director
  structures of defects in Grandjean-Cano wedges of cholesteric liquid crystals
  studied by fluorescence confocal polarizing microscopy}. Phys Rev E,
  66:051703.

\bibitem{Gennes1993}
de~Gennes, P, Prost, J, 1993;\ \emph{The Physics of Liquid Crystals}.
  Oxford Science Pulblications.

\bibitem{Pawsey2012}
Pawsey, AC, Lintuvuori, JS, Wood, TA, Thijssen, JHJ, Marenduzzo, D, Clegg, PS,
  2012;\ {Colloidal particles at the interface between an isotropic
  liquid and a chiral liquid crystal}. Soft Matter, 8422--8428.

\bibitem{Dierking2012}
Dierking, I, Blenkhorn, W, Credland, E, Drake, W, Kociuruba, R, Kayser, B,
  Michael, T, 2012;\ {Stabilising liquid crystalline Blue Phases}.
  Soft Matter, 8, 16:4355.

\bibitem{Rozic2011}
Ro\v{z}i\v{c}, B, Tzitzios, V, Karatairi, E, Tkalec, U, Nounesis, G, Kutnjak,
  Z, Cordoyiannis, G, Rosso, R, Virga, EG, Mu\v{s}evi\v{c}, I, Kralj, S,
  2011;\ {Theoretical and experimental study of the
  nanoparticle-driven blue phase stabilisation.} E Phys J E, 34, 2:1--11.

% \bibitem{Pawsey2014}
% Pawsey, AC, 2014;\ \emph{Colloids at liquid crystal interfaces}.
%   Ph.D. thesis, University of Edinburgh.

\end{thebibliography}
\end{document}